\documentclass[useAMS,fleqn,usenatbib]{mnras}
\usepackage{hyperref}
\usepackage{graphicx}
\usepackage{amssymb} 
\usepackage{amsmath} 
\def\aap{A\&A}

\def\apj{ApJ}
\def\apjs{ApJS}
\def\apjl{ApJL}

\def\mnras{MNRAS}
\def\aj{AJ}
\def\nat{Nature}

\def\prd{Phys. Rev. D}

 \def\la{\mathrel{\mathpalette\fun <}}
 \def\ga{\mathrel{\mathpalette\fun >}}
 \def\fun#1#2{\lower3.6pt\vbox{\baselineskip0pt\lineskip.9pt
        \ialign{$\mathsurround=0pt#1\hfill##\hfil$\crcr#2\crcr\sim\crcr}}}

\title[Combining Gaussian posterior distributions]
{
The clustering of galaxies in the completed SDSS-III Baryon Oscillation
 Spectroscopic Survey: combining correlated Gaussian posterior distributions
}
\author[A. G. S\'anchez et al.]{Ariel G. S\'anchez,$^{1}$\thanks{E-mail: arielsan@mpe.mpg.de}
Jan~Niklas Grieb$^{2,1}$,
Salvador Salazar-Albornoz$^{2,1}$,
Shadab Alam$^{3,4}$,
\newauthor Florian Beutler$^{5,6}$, 
Ashley~J. Ross$^{7,5}$,
Joel~R. Brownstein$^{8}$,
Chia-Hsun Chuang$^{9,10}$,
\newauthor Antonio~J. Cuesta$^{11}$,
Daniel~J. Eisenstein$^{12}$,
Francisco-Shu Kitaura$^{10,6,13}$,
Will~J. Percival$^{5}$,
\newauthor Francisco Prada$^{8,14,15}$,
Sergio Rodr\'{\i}guez-Torres$^{8,14,16}$,
Hee-Jong Seo$^{17}$,
Jeremy Tinker$^{18}$, 
\newauthor Rita Tojeiro$^{19}$,
Mariana Vargas-Maga\~na$^{20,3,4}$,
Jose~A. Vazquez$^{21}$ \&
Gong-Bo Zhao$^{22,5}$
\\
$^{1}$ Max-Planck-Institut f\"ur extraterrestrische Physik, Postfach 1312, Giessenbachstr., 85741 Garching, Germany\\
$^{2}$ Universit\"ats-Sternwarte M\"unchen, Ludwig-Maximilians-Universit\"at M\"unchen, Scheinerstrasse 1, 81679 Munich, Germany\\
$^{3}$ Department of Physics, Carnegie Mellon University, 5000 Forbes Ave., Pittsburgh, PA 15217, USA\\
$^{4}$ McWilliams Center for Cosmology, Carnegie Mellon University, 5000 Forbes Ave., Pittsburgh, PA 15217, USA\\
$^{5}$ Institute of Cosmology \& Gravitation, University of Portsmouth, Dennis Sciama Building, Portsmouth PO1 3FX, UK\\
$^{6}$ Lawrence Berkeley National Laboratory, 1 Cyclotron Road, Berkeley, CA 94720, USA \\
$^{7}$ Center for Cosmology and Astro-Particle Physics, Ohio State University, Columbus, OH 43210, USA\\
$^{8}$ Department of Physics and Astronomy, University of Utah, 115 S 1400 E, Salt Lake City, UT 84112, USA\\
$^{9}$ Instituto de F\'{\i}sica Te\'orica, (UAM/CSIC), Universidad Aut\'onoma de Madrid, Cantoblanco, E-28049 Madrid, Spain\\
$^{10}$ Leibniz-Institut f\"ur Astrophysik Potsdam (AIP), An der Sternwarte 16, D-14482 Potsdam, Germany \\
$^{11}$ Institut de Ci\`encies del Cosmos (ICCUB), Universitat de Barcelona (IEEC-UB), Mart\'i i Franqu\`es 1, E08028 Barcelona, Spain\\
$^{12}$ Harvard-Smithsonian Center for Astrophysics, 60 Garden St., Cambridge, MA 02138, USA \\
$^{13}$ Departments of Physics and Astronomy, University of California, Berkeley, CA 94720, USA \\
$^{14}$ Campus of International Excellence UAM+CSIC, Cantoblanco, E-28049 Madrid, Spain\\
$^{15}$ Instituto de Astrof\'{\i}sica de Andaluc\'{\i}a (CSIC), Glorieta de la Astronom\'{\i}a, E-18080 Granada, Spain\\
$^{16}$ Departamento de F\'{\i}sica Te\'orica, Universidad Aut\'onoma de Madrid, Cantoblanco, 28049, Madrid, Spain\\
$^{17}$ Department of Physics and Astronomy, Ohio University, 251B Clippinger Labs, Athens, OH 45701, USA\\
$^{18}$ Center for Cosmology and Particle Physics, New York University, New York, NY 10003, USA\\
$^{19}$ School of Physics and Astronomy, University of St Andrews, North Haugh, St Andrews KY16 9SS, UK\\
$^{20}$ Instituto de Fisica, Universidad Nacional Aut\'onoma de M\'exico, Apdo. Postal 20-364, M\'exico\\
$^{21}$ Brookhaven National Laboratory, Bldg 510, Upton, New York 11973, USA \\
$^{22}$ National Astronomy Observatories, Chinese Academy of Science, Beijing, 100012, P.R.China
}

\date{Submitted to MNRAS}
\pubyear{2016}

\begin{document}
\label{firstpage}
\pagerange{\pageref{firstpage}--\pageref{lastpage}}
\maketitle

\begin{abstract}
The cosmological information contained in anisotropic galaxy clustering measurements can often be
compressed into a small number of parameters whose posterior distribution is well
described by a Gaussian. We present a general methodology to combine these estimates into a single 
set of consensus constraints that encode the total information of the individual measurements, 
taking into account the full covariance between the different methods. 
We illustrate this technique by applying it to combine the results obtained from different 
clustering analyses, including measurements of the signature of baryon acoustic oscillations (BAO) 
and redshift-space distortions (RSD), based on a set of mock catalogues of the final 
SDSS-III Baryon Oscillation Spectroscopic Survey (BOSS).
Our results show that the region of the parameter space allowed by the consensus constraints is
smaller than that of the individual methods, highlighting the importance of performing multiple analyses
on galaxy surveys even when the measurements are highly correlated.
This paper is part of a set that analyses the final galaxy clustering dataset from BOSS.
The methodology presented here is used in \citet{Acacia2016} to produce the final 
cosmological constraints from BOSS.
\end{abstract}

\section{Introduction}
\label{sec:intro}

Over the past decades the size and quality of galaxy redshift surveys has increased dramatically. 
Thanks to these data sets, the information from the large-scale structure (LSS) 
of the Universe has played a central role in
establishing the current cosmological paradigm,
the $\Lambda$CDM model \citep[e.g.][]{Tegmark2004,
Eisenstein2005, Cole2005, Anderson2012, Anderson2013, Anderson2014}. 

Several methods can be used to extract the information encoded in the large-scale distribution of
galaxies. The power spectrum, $P(k)$,
and its Fourier transform, the two-point correlation function $\xi(s)$, have been
the preferred tools for LSS analyses. The anisotropies in these measurements caused by redshift-space
distortions (RSD) and the Alcock--Paczynski effect \citep{Alcock1979}
can be studied by means of their Legendre multipoles \citep[e.g.][]{Padmanabhan2008} 
or using the clustering wedges statistic \citep{Kazin2012}.
Thanks to the combined information of baryon acoustic oscillations (BAO) and 
RSD, anisotropic clustering measurements
can simultaneously constrain the expansion history of the Universe and the growth of density
fluctuations, thus offering one of the most powerful cosmological probes.

The potential of LSS observations as cosmological probes has led to the construction of increasingly
larger galaxy catalogues. Examples of these new surveys include the completed Baryon Oscillation Spectroscopic Survey 
\citep[BOSS;][]{Dawson2013}, which is part of the Sloan Digital Sky Survey III \citep[SDSS-III;][]{Eisenstein2011}, the 
on-going SDSS-IV extended Baryon Oscillation Spectroscopic Survey \citep[eBOSS;][]{Dawson2016}
and future surveys such as the Hobby Eberly Telescope Dark Energy Experiment \citep[HETDEX;][]{Hill:2008mv}, 
the Dark Energy Spectroscopic Instrument \citep[DESI;][]{Levi:2013gra} and the ESA space mission \emph{Euclid} \citep{Laureijs:2011gra}.

As the construction of galaxy surveys requires a considerable amount of resources from the community, 
substantial efforts are put into maximizing the information extracted from the obtained data sets.
This problem has often been posed as that of determining which statistic is the best to 
extract cosmological information (e.g. power spectrum vs. correlation function), often based on a 
simple metric or figure of merit. However, although the results obtained by applying 
different statistics to a given data set are highly correlated, as they are
based on estimators and each measurement is analysed over a limited range of scales, they do not
contain exactly the same information or are affected by noise in the same way. This means that, if the 
covariance between the different measurements is correctly taken into account, additional information could 
be obtained by combining the results inferred from different methods.

In most cases, the cosmological information contained in the clustering measurements can be
condensed into a small number of parameters whose posterior distribution is well
described by a multivariate Gaussian. In this case, the obtained constraints can be represented by the mean values of these
parameters and their respective covariance matrices.  
The analyses of the final BOSS galaxy samples of our companion papers are examples of 
this situation \citep{Beutler2016a, Beutler2016b, Grieb2016, Ross2016, Sanchez2016, Satpathy2016}.
The BAO and RSD information obtained in these analyses can be expressed as constraints on  
the ratio of the comoving angular diameter distance 
to the sound horizon at the drag redshift, $D_{\rm M}(z)/r_{\rm d}$, the product of the 
Hubble parameter and the sound horizon, $H(z)\times r_{\rm d}$, and 
the growth-rate of cosmic structures, characterized by the combination $f\sigma_8(z)$, where 
$f(z)$ is the logarithmic growth rate and $\sigma_8(z)$ represents the linear rms mass fluctuation
in spheres of radius $8\,h^{-1}{\rm Mpc}$. 

Here we present a general methodology to combine several Gaussian posterior distributions 
into a single set of consensus constraints representing their joint information,  
taking into account the full covariance between the different estimates. 
We illustrate this technique by applying it to the results
inferred from the application of the same clustering analyses performed on the final BOSS galaxy samples 
to 996 {\sc Multidark-Patchy} ({\sc MD-Patchy}) mock galaxy catalogues reproducing the properties of the
survey \citep{Kitaura2016}.
The obtained consensus distributions represent a gain in constraining power with respect to the results of
the individual methods, highlighting the importance of performing multiple analyses on galaxy surveys.
The methodology presented here is used in our companion paper \citet{Acacia2016} to combine the 
cosmological information from the different analyses methods applied to the final BOSS galaxy samples 
\citep{Beutler2016a,Beutler2016b,Grieb2016,Ross2016,Sanchez2016,Satpathy2016}
into a final set of consensus constraints.

The structure of the paper is as follows, in Section \ref{sec:combination} we present the general scheme
for the combination of different Gaussian posterior distributions into a set of consensus constraints that
encode the full information provided by these estimates. We consider the cases in which the posterior
distributions cover the same parameter spaces and when they differ. In Section~\ref{sec:app}
we illustrate this procedure by applying it to the results obtained from different 
BAO and RSD measurements from a set of BOSS mock catalogues. 
Finally, Section~\ref{sec:conclusions} contains our main conclusions.

\section{The combination of Gaussian posterior distributions}
\label{sec:combination}

In this section we describe the general formalism to combine the information from several 
posterior distributions into a set of consensus constraints that fully account for their
covariance. We begin with the case in which all distributions contain the same parameters
and later extend these results to the more general case in which the overlap can be partial.

\subsection{The combination of posterior distributions on the same parameter space}

Let us assume that $m$ different statistical analyses have been performed on a given data set, 
each of them producing an estimate of the same set of $p$ parameters.  
If the posterior distributions of these parameters are well described by a Gaussian, 
the results of any given method $i$ can be represented by an array of $p$ measurements
${\bf D}_i$ and their corresponding $p\times p$ covariance matrix $\mathbfss{C}_{ii}$. 
Considering all $m$ methods, the full set of measurements can be written in a single array
of dimension $m\cdot p$  as 
\begin{equation}
{\bf D}_{\rm tot} = ({\bf D}_1, \cdots, {\bf D}_{m}),
\end{equation}
with a total 
covariance matrix
\begin{equation}
\mathbfss{C}_{\rm tot} = \left( \begin{array}{ccc}
\mathbfss{C}_{11} & \cdots & \mathbfss{C}_{1m} \\
\vdots & \ddots & \vdots \\
\mathbfss{C}_{m1} & \cdots & \mathbfss{C}_{mm} \\ \end{array} \right),
\end{equation}
where each block $\mathbfss{C}_{ij}$ represents the cross-covariance matrix between the results 
of methods $i$ and $j$.  

\begin{figure*}
\includegraphics[width=0.98\textwidth]{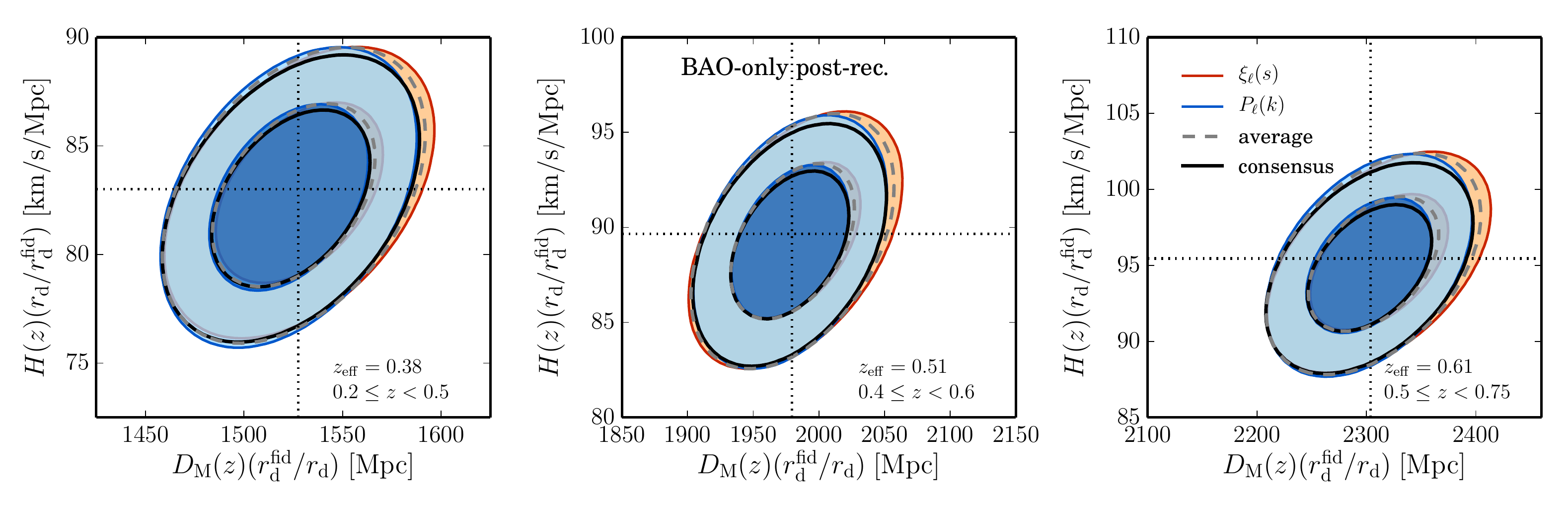}
\caption{
The mean 68\% and 95\% two-dimensional constraints on the parameters $D_{\rm M}(z)(r_{\rm d}^{\rm fid}/r_{\rm d})$ and
$H(z)(r_{\rm d}/r_{\rm d}^{\rm fid})$ obtained by applying the BAO-only analyses of \citet[orange]{Ross2016} and 
\citet[blue]{Beutler2016a} to 996 {\sc MD-Patchy} BOSS mock catalogues for the redshift bins indicated
in the legend.
The results are in excellent agreement with the true underlying values of these parameters, indicated by the dotted lines.
The full information from these measurements can be combined into a set of consensus constraints (black solid lines) 
as described in Section~\ref{sec:combination}. The dashed lines correspond to the combination of the results obtained by averaging the logarithm of the two posterior distributions.
}
\label{fig:2d_bao}
\end{figure*}

A given model will predict values for these parameters, which we will represent by the array ${\bf T}$.
Defining
\begin{equation}
{\bf  T}_{\rm tot} = ({\bf T}, \cdots, {\bf T}),
\end{equation}
that is, ${\bf T}$ repeated $m$ times, and a total precision matrix as
\begin{equation}
\mathbf{\Psi}_{\rm tot} \equiv \mathbfss{C}_{\rm tot}^{-1}
\end{equation}
we can compute the $\chi^2$ of a model taking into account the combined information of 
all measurements as
\begin{equation}
\chi^2 = ({\bf D}_{\rm tot}-{\bf T}_{\rm tot})^{\rm t}\mathbf{\Psi}_{\rm tot}({\bf D}_{\rm tot}-{\bf T}_{\rm tot}).
\label{eq:chi2_total}
\end{equation}

Our goal is to compress the combined information of all the measurements into a single set of $p$ 
consensus values, ${\bf D}_{\rm c}$, with its corresponding $ p\times p$
covariance matrix, $\mathbfss{C}_{\rm c}$, such that
\begin{equation}
\chi_{\rm c}^2 = ({\bf D}_{\rm c}-{\bf T})^{\rm t}{\bf \Psi}_{\rm c}({\bf D}_{\rm c}-{\bf T}),
\label{eq:chi2_comb}
\end{equation}
where
\begin{equation}
{\bf \Psi}_{\rm c}=\mathbfss{C}_{\rm c}^{-1},
\end{equation}
is equal to the $\chi^2$ value of equation (\ref{eq:chi2_total}) up to an additive constant, which would
only correspond to a re-normalization of the likelihood function.
In order to do this we first write the full precision matrix, ${\bf \Psi}_{\rm tot}$, in blocks 
of size $p\times p$ as 
\begin{equation}
{\bf \Psi}_{\rm tot} = \left( \begin{array}{ccc}
{\bf \Psi}_{11} & \cdots & {\bf \Psi}_{1m} \\
\vdots & \ddots & \vdots \\
{\bf \Psi}_{m1} & \cdots & {\bf \Psi}_{mm} \\ \end{array} \right).
\end{equation}
Note that, in general, ${\bf \Psi}_{ij}$ is not the inverse of the corresponding block $\mathbfss{C}_{ij}$ in $\mathbfss{C}_{\rm tot}$.

The solution for ${\bf D}_{\rm c}$ and $\mathbfss{C}_{\rm c}$ can be found easily by expanding the
expression for the total $\chi^2$ of equation (\ref{eq:chi2_total}) as
\begin{equation}
\chi^2 = {\bf D}_{\rm tot}^{\rm t}\mathbf{\Psi}_{\rm tot}{\bf D}_{\rm tot}-2\,{\bf T}_{\rm tot}^{\rm t}\mathbf{\Psi}_{\rm tot}{\bf D}_{\rm tot} + {\bf T}_{\rm tot}^{\rm t}\mathbf{\Psi}_{\rm tot}{\bf T}_{\rm tot}. 
\label{eq:chi2_total_expanded}
\end{equation}
Equivalently, for the consensus values we will have
\begin{equation}
\chi^2 = {\bf D}_{\rm c}^{\rm t}\mathbf{\Psi}_{\rm c}{\bf D}_{\rm c}-2\,{\bf T}^{\rm t}\mathbf{\Psi}_{\rm c}{\bf D}_{\rm c} + {\bf T}^{\rm t}\mathbf{\Psi}_{\rm c}{\bf T}. 
\label{eq:chi2_cons_expanded}
\end{equation}
Equating the last terms of equations (\ref{eq:chi2_total_expanded}) and (\ref{eq:chi2_cons_expanded}),
we find a general expression for $\mathbfss{C}_{\rm c}$ as
\begin{equation}
\mathbfss{C}_{\rm c} \equiv \Psi_{\rm c}^{-1} \equiv \left(\sum_{i=1}^{m}\sum_{j=1}^{m} \Psi_{ij}\right)^{-1},
\label{eq:C_comb}
\end{equation}
while equating the second terms gives the solution for ${\bf D}_{\rm c}$ as
\begin{equation}
{\bf D}_{\rm c} = \mathbf{\Psi}_{\rm c}^{-1} \sum_{i=1}^{m}\left(\sum_{j=1}^{m} \Psi_{ji}\right){\bf D}_i.
\label{eq:D_comb}
\end{equation}

It is easy to see that in the case in which the different estimates are independent, these
expressions reduce to the known formulae
\begin{equation}
\Psi_{\rm c} = \sum_{i=1}^{m} \Psi_{ii},
\end{equation}
and
\begin{equation}
{\bf D}_{\rm c} = \Psi_{\rm c}^{-1} \sum_{i=1}^{m} \Psi_{ii}{\bf D}_i,
\end{equation}
where ${\bf \Psi}_{ii}$ corresponds to the precision matrix of measurement $i$. 

Another interesting particular case is when the goal is to obtain the consensus value of a single
parameter (i.e. $p=1$) given a set of $m$ measurements $D_i$. In this case, equations (\ref{eq:C_comb})
and (\ref{eq:D_comb}) show that the consensus mean and dispersion for this parameter will be 
given by
\begin{equation}
D_{\rm c} = \sigma_{\rm c}^2 \sum_{i=1}^m\sum_{j=1}^m \psi_{ij}D_i,
\end{equation}
and
\begin{equation}
\sigma_{\rm c}^2 = \left(\sum_{i=1}^{m}\sum_{j=1}^{m} \psi_{ij}\right)^{-1},
\end{equation}
which correspond to the result found by \citet{Wrinkler1981}. 

\subsection{The combination of posterior distributions with different parameters}
\label{sec:combination_dif}

In certain cases, it might be necessary to combine two or more posterior 
distributions with different parameters. This situation is encountered, 
for example, when combining cosmological distance measurements obtained from BAO-only 
analyses with the information obtained from full-shape fits to anisotropic clustering
measurements, which also constrain
the growth-rate parameter combination $f\sigma_8(z)$.

The recipe described in the previous section can also be applied in this case. 
As an example, let us consider the case in which the first data set gives constraints
on the first $p-1$ parameters only, with an associated 
$(p-1)\times(p-1)$ covariance matrix $\tilde{\mathbfss{C}}_{11}$ .
These results can be considered as including a constraint on the remaining parameter,
but with an infinite uncertainty, that is 
\begin{equation}
\mathbfss{C}_{11}=\left( \begin{array}{cc}
\tilde{\mathbfss{C}}_{11} & 0 \\
0 & \infty \\ \end{array} \right).
\end{equation}
In the remaining blocks of the total covariance matrix $\mathbfss{C}_{\rm tot}$ the rows and
columns corresponding to the undetermined parameter will be zero. This structure will  
be inherited by the total precision matrix, where also the diagonal entry corresponding to this
parameter will cancel.
It is then possible to apply the solution of equations (\ref{eq:C_comb}) and (\ref{eq:D_comb})
to find the final consensus values that combine all measurements.  

In a more general situation, given a set of measurements of different parameter
spaces, it is possible to apply the general recipe described here to obtain consensus values
on the parameter space defined by the union of those of the individual measurements.

\section{Application to BAO and RSD measurements from BOSS}
\label{sec:app}

As an illustration of the procedure described in the previous section
we have applied the procedure described in the previous section to assess the combination 
of the information obtained from a set of cosmological measurements made on mock catalogues designed 
to mimic the BOSS DR12 sample. The MD-Patchy mock galaxy catalogues \citep{Kitaura2016}, of which we 
use 996, are based on a cosmological model corresponding to the best fitting CDM cosmology to the 
Planck 2013 CMB measurements \citep{PlanckXVI2013}. We followed \citet{Acacia2016} and divided each 
mock catalogue into three overlapping redshift bins of roughly equal volume, defined by 
$0.2 < z < 0.5$, $0.4 < z < 0.6$ and $0.5 < z < 0.75$.
We focus first on the combination of the results of the BAO-only and full-shape fits separately
and then combine these constraints into a final set of consensus values.

\subsection{Post-reconstruction BAO-only fits}
\label{sec:bao}

For each {\sc MD-Patchy} mock catalogue we applied the  
methodologies of \citet{Ross2016} and \citet{Beutler2016a} to 
perform BAO-only fits to the Legendre multipoles of order $\ell=0,2$ of the two-point functions
in configuration and Fourier space obtained in each of our three redshift bins after the 
application of the reconstruction technique \citep{Eisenstein2007b,Padmanabhan2012} as described
in \citet{Cuesta2016}.

\begin{figure}
\centering
\includegraphics[width=0.4\textwidth]{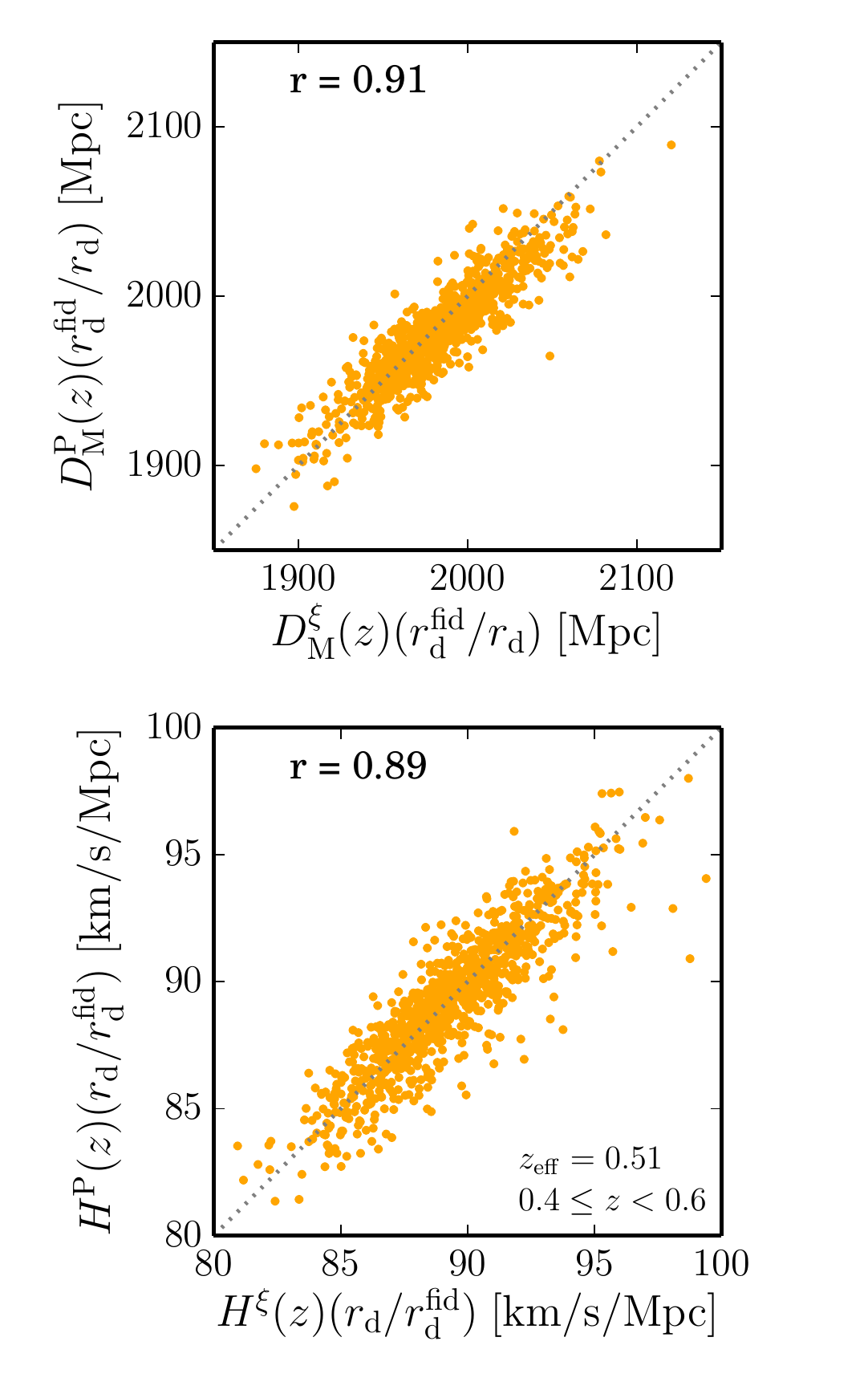}
\caption{
Scatter plots of the BAO-only constraints on $D_{\rm M}(z)(r_{\rm d}^{\rm fid}/r_{\rm d})$ and
$H(z)(r_{\rm d}/r_{\rm d}^{\rm fid})$ obtained from the configuration and Fourier space BAO-only 
analyses of 996 {\sc MD-Patchy} mock catalogues for $0.4 < z < 0.6$. 
Although the results obtained from these methods are highly correlated,
their correlation coefficients, $r$, are not exactly one, indicating that 
additional information can be obtained from their combination.
}
\label{fig:scatter_bao}
\end{figure}

The cosmological information encoded in the BAO signal can be expressed in terms of the 
geometric parameters $D_{\rm M}(z)/r_{\rm d}$, and $H(z)\times r_{\rm d}$.
Figure~\ref{fig:2d_bao} shows the mean two-dimensional constraints on these
parameters for each redshift bin, rescaled by the sound horizon at the drag redshift for
our fiducial cosmology, $r_{\rm d}^{\rm fid}=147.78\,{\rm Mpc}$, to express them in units of Mpc and 
${\rm km}\,{\rm s}^{-1}{\rm Mpc}^{-1}$.
The results inferred from the two methods are completely consistent and in excellent
agreement with the true underlying values of these parameters, which are shown by the dotted lines.
This indicates that both methods are able to extract essentially the same information from 
the clustering measurements.
However, the results obtained from each set of measurements on individual mock catalogues are
affected by noise in different ways. This can be seen in  
Figure~\ref{fig:scatter_bao}, which shows scatter plots of the two sets of constraints 
obtained from each mock catalogue for our intermediate redshift bin.
Although they are highly correlated, the correlation coefficients between the results 
derived from the two methods are not exactly
one, which means that additional information can be obtained by combining them.

\begin{figure}
\includegraphics[width=0.5\textwidth]{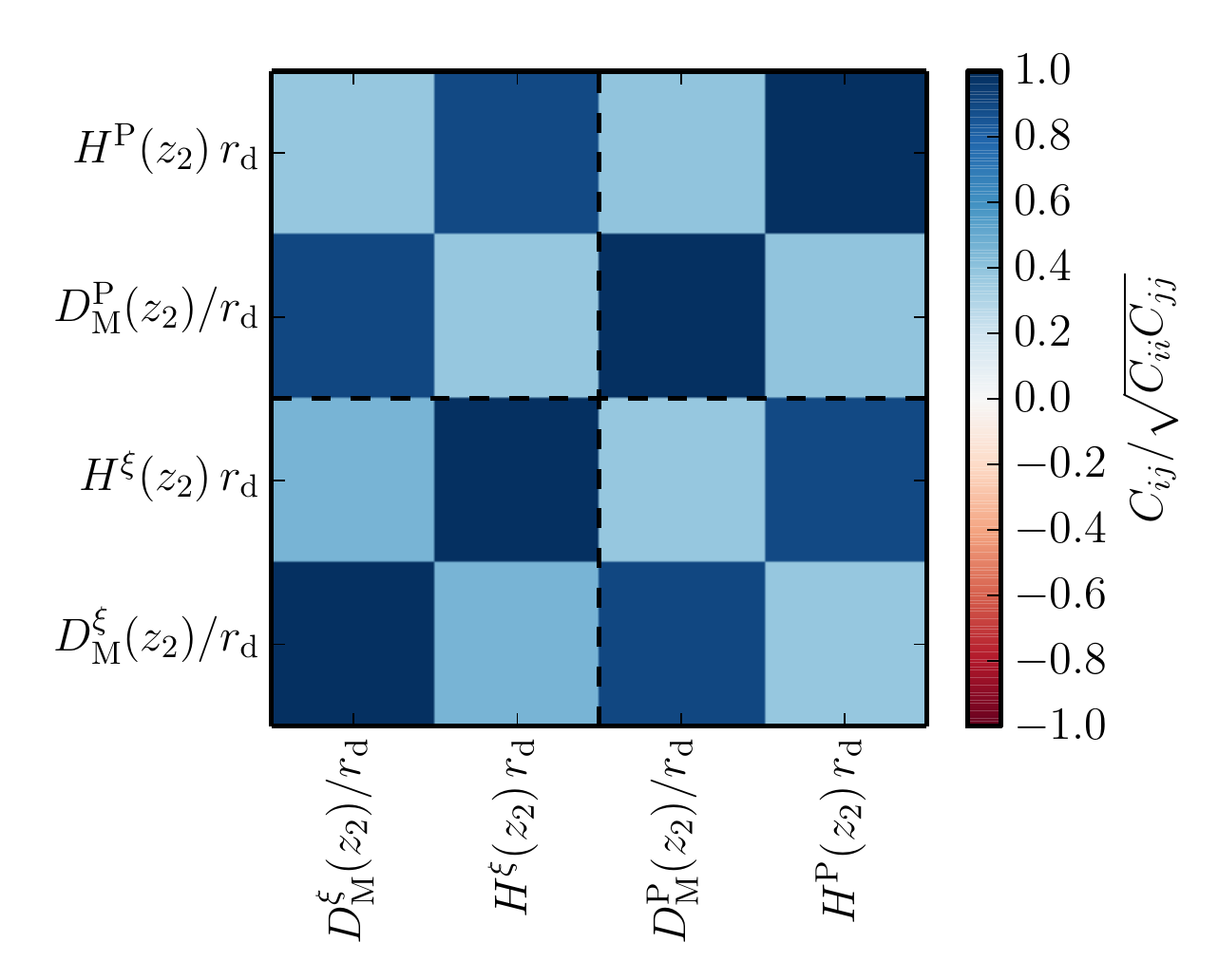}
\caption{
Correlation matrix corresponding to the full covariance $\mathbfss{C}_{\rm tot}$
of the BAO-only constraints on $D_{\rm M}(z)/r_{\rm d}$, and $H(z)\times r_{\rm d}$, constructed
from the individual {\sc MD-Patchy} mock catalogues  in configuration and Fourier space. 
The blocks $\mathbfss{C}_{ij}$ indicated by the dashed lines correspond to the auto and cross-covaraince
matrices of the two methods.
}
\label{fig:fullcov_bao}
\end{figure}

The results obtained from the two methods on each individual mock catalogue can be used to 
construct the total covariance matrix $\mathbfss{C}_{\rm tot}$. As an example, 
Figure~\ref{fig:fullcov_bao} shows the normalized correlation matrix corresponding to the 
results of the intermediate redshift bin. The dashed lines divide the matrix into the
blocks associated with $\mathbfss{C}_{ij}$. Due to the high correlation between the results
of the power spectrum and correlation function fits, the structure of the off-diagonal block 
$\mathbfss{C}_{12}$ is very similar to that of the auto-covariances. 
Inverting the matrix $\mathbfss{C}_{\rm tot}$ to obtain the total precision matrix 
$\mathbf{\Psi}_{\rm tot}$ and using equations~(\ref{eq:C_comb}) and (\ref{eq:D_comb}), the
results of both methods can be combined into sets of consensus constraints for each redshift
bin, which are shown by the black solid lines in Figure~\ref{fig:2d_bao}. As described in 
Section~\ref{sec:combination}, these constraints contain the joint information of the two sets 
of results. 

\begin{figure*}
\includegraphics[width=0.95\textwidth]{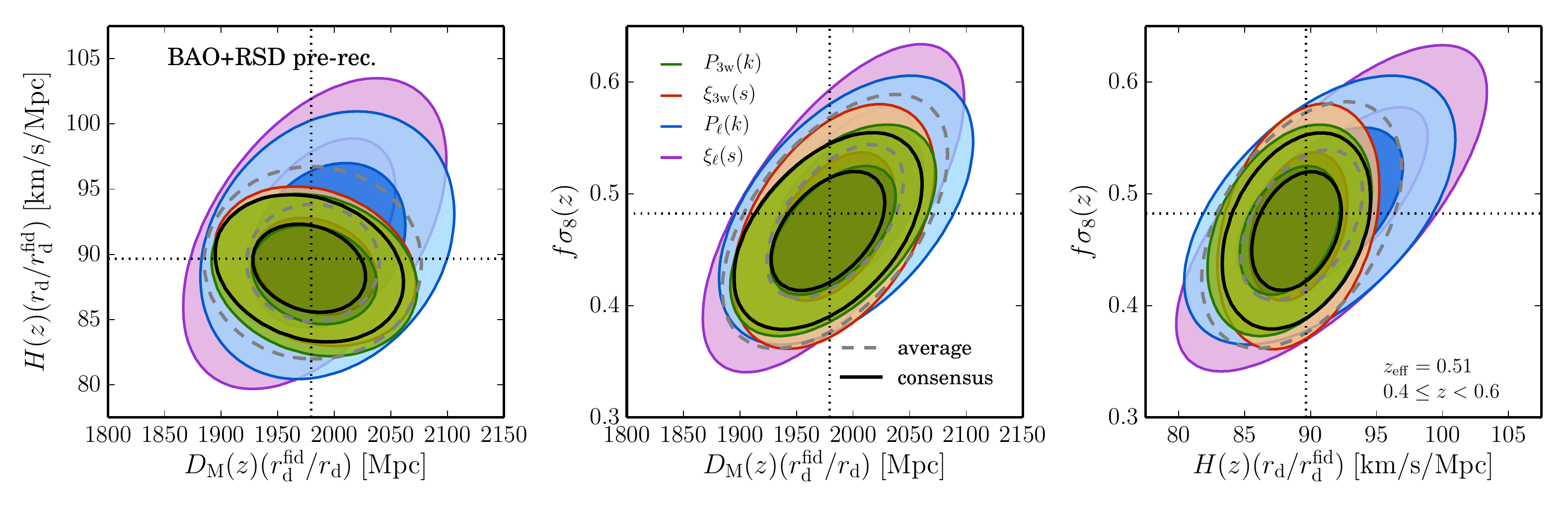}
\caption{
The mean 68\% and 95\% two-dimensional constraints on the parameters $D_{\rm M}(z)(r_{\rm d}^{\rm fid}/r_{\rm d})$, 
$H(z)(r_{\rm d}/r_{\rm d}^{\rm fid})$ and $f\sigma_8(z)$ inferred from our mock BOSS catalogues for 
$0.4 < z < 0.6$. The filled contours correspond to the results obtained by means of 
full-shape fits of the Legendre multipoles, $\xi_{\ell}(s)$ (magenta) and $P_{\ell}(k)$ (blue) and clustering wedges
$\xi_{3{\rm w}}(s)$ (orange) and $P_{3{\rm w}}(k)$ (green), using the methodology of our companion papers \citep{Satpathy2016, Beutler2016b, Sanchez2016, Grieb2016}. The obtained constraints are good agreement with the true underlying values of these parameters,
indicated by the dotted lines.
The black solid contours correspond to the combination of these measurements into a set of consensus constraints,
computed as described in Section~\ref{sec:combination}. The dashed lines correspond to the combination of the results obtained by averaging the logarithms of the four posterior distributions.}
\label{fig:2drsd}
\end{figure*}

\begin{figure}
\includegraphics[width=0.5\textwidth]{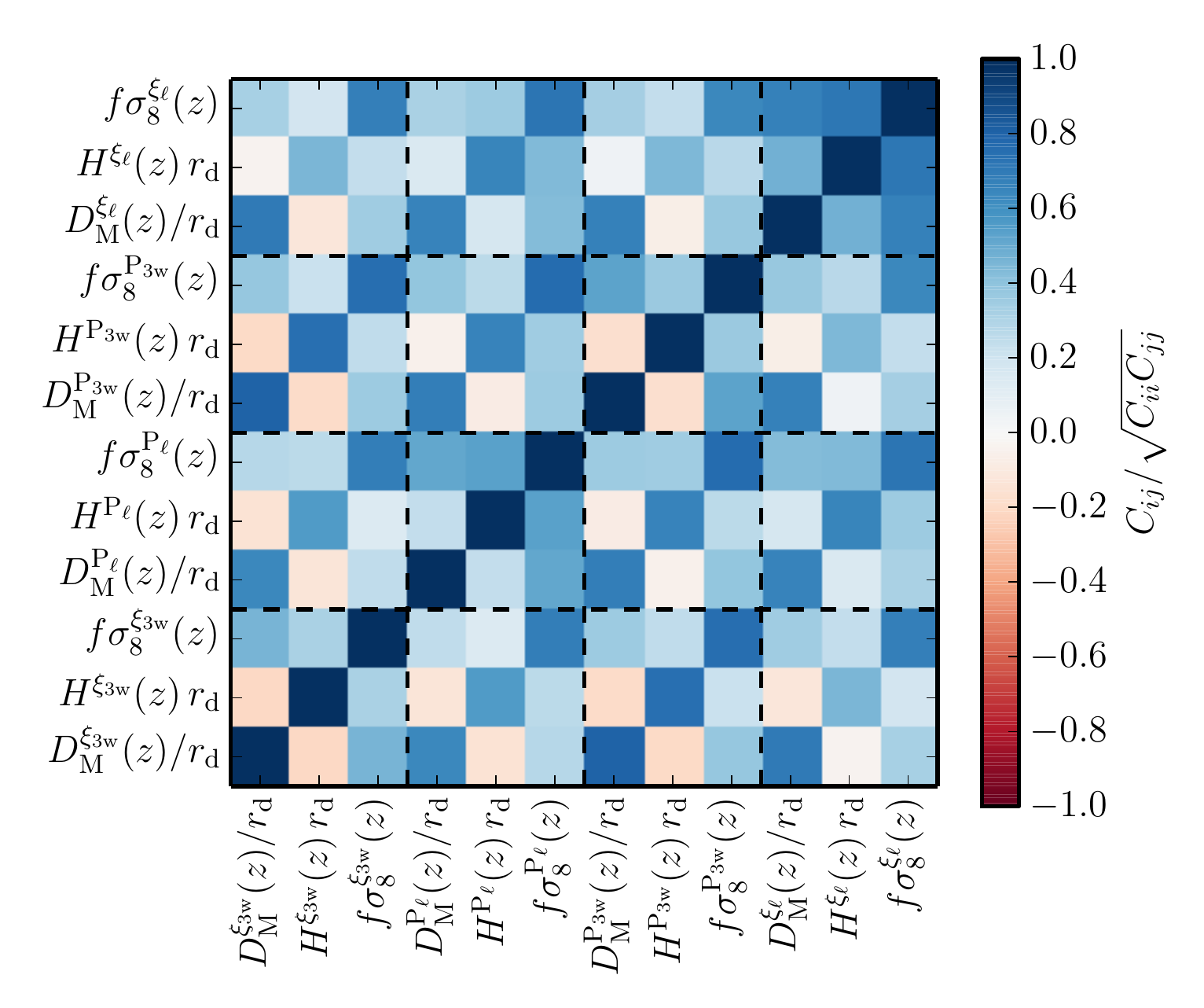}
\caption{
Correlation matrix corresponding to the total covariance $\mathbfss{C}_{\rm tot}$
of the full-shape fits of the Legendre multipoles and clustering wedges in configuration and Fourier space
constructed from the individual {\sc MD-Patchy} mock catalogues. 
The blocks $\mathbfss{C}_{ij}$ indicated by the dashed lines correspond to the auto and cross-covaraince
matrices of the different methods.
}
\label{fig:fullcov_rsd}
\end{figure}

\citet{Anderson2012,Anderson2013,Anderson2014} derived consensus anisotropic BAO constraints from the 
combination of the results inferred from the analysis of the Legendre multipoles of the correlation
function and clustering wedges statistic \citep*{Kazin2012}. These constraints were 
computed by averaging the logarithm of the posterior distributions obtained from each method.
The dashed lines in Figure~\ref{fig:2d_bao} show the result of applying this procedure to the
constraints inferred from the Fourier and configuration-space fits to our mock catalogues. 
As the original distributions are similar, their average is also in agreement with the 
full consensus constraints, but results in a slightly larger allowed region for     
$D_{\rm M}(z)/r_{\rm d}$ and $H(z)\times r_{\rm d}$. This difference can be quantified 
in terms of the Figure of merit, $FoM$, given by
\begin{equation}
FoM=\left(\det[\mathbfss{C}]\right)^{-\frac{1}{2}}
\label{eq:fom}
\end{equation}
The $FoM$ values of the consensus constraints are larger than those of the 
average profile by a factor 1.07, 1.08 and 1.10 for the low-, intermediate- and 
high-redshift bins. As we will see in the next sections, 
this is a common feature of the result of the average profile. 

\begin{figure*}
\includegraphics[width=0.95\textwidth]{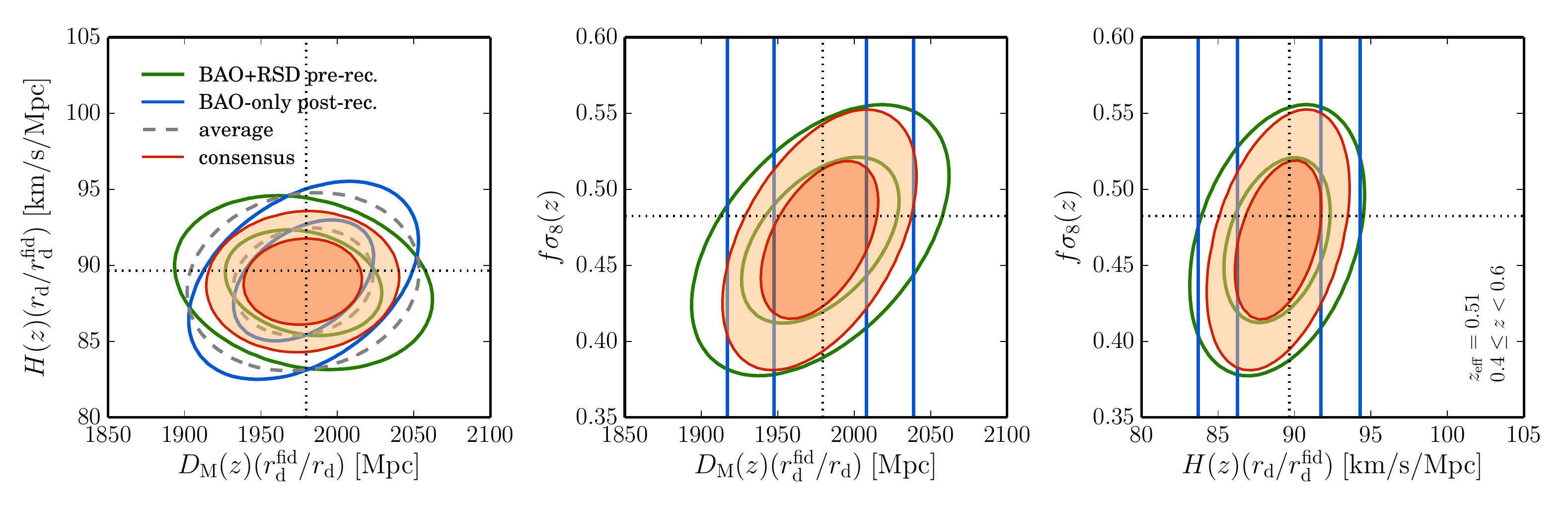}
\caption{
The mean 68\% and 95\% two-dimensional consensus constraints on the parameters 
$D_{\rm M}(z)(r_{\rm d}^{\rm fid}/r_{\rm d})$, $H(z)(r_{\rm d}/r_{\rm d}^{\rm fid})$ and $f\sigma_8(z)$ 
inferred from our mock BOSS catalogues for $0.4 < z < 0.6$. 
The blue and green contours correspond to the combination of the BAO-only and 
full-shape BAO+RSD fits, respectively. The dotted lines indicate the correct values of these parameters.
The filled contours correspond to the combination of these results into a final set of consensus constraints,
containing the joint information of the two sets of measurements. The dashed contours in the left panel
correspond to the result obtained by averaging the logarithms of the two posterior distributions.
}
\label{fig:2d_final}
\end{figure*}

\subsection{Pre-reconstruction full-shape fits}
\label{sec:rsd}

In this Section we focus on the combination of the results inferred from full-shape fits 
(to which we refer as BAO+RSD analyses) to various pre-reconstruction anisotropic clustering
measurements.
We consider the analysis methods applied to the final BOSS data in our companion papers
\citep{Beutler2016b, Grieb2016, Sanchez2016, Satpathy2016}, to constrain the same geometric
parameters as the BAO-only studies, $D_{\rm M}(z)/r_{\rm d}$ and $H(z)\times r_{\rm d}$,
and the growth rate of cosmic structure, characterized by the combination $f\sigma_8(z)$, where
\begin{equation}
f = \frac{{\rm d}\ln D}{{\rm d}\ln a}
\end{equation}
is the logarithmic derivative of the growth factor. 
\citet{Satpathy2016} use a model based on convolution Lagrangian perturbation theory 
\citep[CLPT;][]{Carlson2013,Wang13} and the Gaussian streaming model \citep{Scoccimarro2004, Reid2011} to fit the
full shape of the monopole and quadrupole of the two-point correlation function, $\xi_{0,2}(s)$. 
\citet{Beutler2016b} apply a model based on \citet{Taruya2010} to the power spectrum multipoles
$P_{\ell}(k)$ for $\ell=0,2,4$. \citet{Grieb2016} and \citet{Sanchez2016} use a new model of the non-linear
evolution of density fluctuations (gRPT; Crocce, Blas \& Scoccimarro in prep.) and RSD to extract cosmological information from
the full shape of three clustering wedges measured in Fourier and configuration space, 
$P_{3{\rm w}}(k)$ and $\xi_{3{\rm w}}(s)$, respectively. 
We performed the same analyses to each of our mock catalogues in the same
way as they were applied to the real BOSS data.

Figure~\ref{fig:2drsd} shows the mean 68\% and 95\% C.L. constraints on 
$D_{\rm M}(z)/r_{\rm d}$, $H(z)\times r_{\rm d}$ and $f\sigma_8(z)$ 
of the results inferred from each individual mock catalogue for our intermediate redshift bin.
The filled contours correspond to the results obtained from the correlation function
multipoles $\xi_{0,2}(s)$ (magenta), the power spectrum multipoles $P_{0,2,4}(k)$ (blue), the correlation
function wedges $\xi_{3{\rm w}}(s)$ (orange) and the power spectrum wedges 
$P_{3{\rm w}}(k)$ (green). The results obtained from these measurements are completely consistent 
and in good agreement with the correct values for the cosmology of the mock catalogues, 
shown by the dotted lines. However, the different measurements and range of scales 
included in each analysis, as well as the models applied to these data, lead to results with larger
differences than in the BAO-only case. We used the methodology described in Section~\ref{sec:combination} to 
obtain a set of consensus values representing the joint information from these analyses. 

We used the results obtained from the application of the different methods
to our mock catalogues to construct the full covariance matrices $\mathbfss{C}_{\rm tot}$ 
in our three redshift bins. 
As an example, Figure~\ref{fig:fullcov_rsd} shows the corresponding correlation matrix for the intermediate 
redshift bin. The dashed lines divide the matrix into the blocks $\mathbfss{C}_{ij}$, corresponding to the
auto and cross-covariance matrices of the methods. 
The different estimates of each parameter are highly correlated. 
The differences between the methods are also reflected in the structure of the
correlation matrix, which is more complicated than for the BAO-only case.

We used equations (\ref{eq:C_comb}) and (\ref{eq:D_comb}) to derive consensus constraints for each mock catalogue.
The back solid contours in Figure~\ref{fig:2drsd} correspond to the mean consensus constraints.
As the consensus results combine the information of all four measurements, 
they provide tighter constraints than each of them individually. 
This highlights the gain obtained from the combination of the methods, with respect to the individual analyses.  

The grey dashed lines in the same figure correspond to the results obtained by averaging the logarithm of the 
posterior distributions recovered from the different methods. The difference between these constraints and
the consensus values can be quantified by extending the definition of the $FoM$ from equation~(\ref{eq:fom}) 
to the three-dimensional covariance matrices of the consensus and average constraints.
In this case the $FoM$ values of the consensus constraints are larger than those of the average 
profile by a factor $\backsim 2.5$ in all redshift bins. 
This difference clearly shows that the average profile does not reproduce the full
information of the different estimates. 

\subsection{Final consensus constraints}
\label{sec:full}

In this section we focus on the combination of the consensus BAO-only constraints derived
in Section~\ref{sec:bao}, 
which are only sensitive to the geometric quantities $D_{\rm M}(z)/r_{\rm d}$ and $H(z)\times r_{\rm d}$, 
with those of the full-shape BAO+RSD measurements derived in Section~\ref{sec:rsd}, which also include
$f\sigma_8(z)$. As these posterior distributions contain different 
parameters, we proceed as described in Section~\ref{sec:combination_dif} and interpret the 
BAO-only results as providing an estimate of $f\sigma_8(z)$ with infinite uncertainty.
The blue and green contours of Figure~\ref{fig:2d_final} show the consensus constraints on our intermediate
redshift bin for the BAO-only and BAO+RSD cases, respectively. As can be seen in the left panel, the
constraints in the $D_{\rm M}(z)/r_{\rm d}$ -- $H(z)\times r_{\rm d}$ plane follow different correlations,
which suggests that their combination could lead to a significant improvement of the constraints.

\begin{figure}
\includegraphics[width=0.45\textwidth]{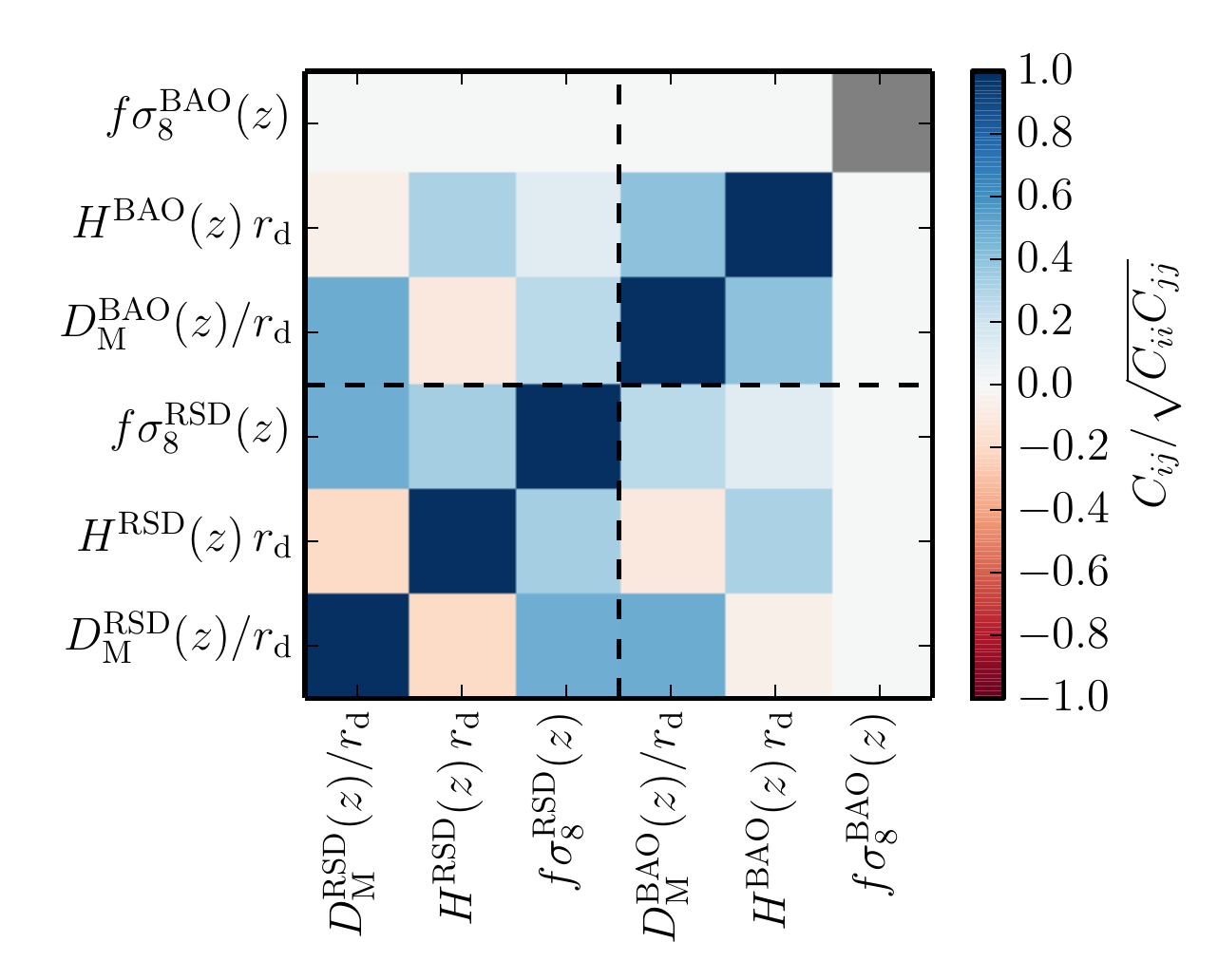}
\caption{
Correlation matrix corresponding to the joint covariance $\mathbfss{C}_{\rm tot}$
of the BAO-only and full-shape consensus results derived in Sections~\ref{sec:bao} and \ref{sec:rsd}. 
The dashed lines indicate the blocks $\mathbfss{C}_{ij}$ corresponding to the auto and cross-covariance
matrices of the two methods.
As BAO-only measurements cannot constraint the value of $f\sigma_8(z)$, the
corresponding diagonal entry is undetermined (shown in grey) and 
its row and column are set to zero.
}
\label{fig:fullcov_final}
\end{figure}

We used the BAO-only and BAO+RSD consensus values inferred from each mock catalogue in Sections~\ref{sec:bao} and \ref{sec:rsd} to obtain the covariance matrix $\mathbfss{C}_{\rm tot}$ associated with these constraints. 
Figure~\ref{fig:fullcov_final} shows the associated correlation matrix, where the
diagonal entry corresponding to the BAO-only estimate of $f\sigma_8(z)$ is undetermined (shown in grey) and 
its corresponding row and column are set to zero. This structure is repeated in the 
total precision matrix $\mathbf{\Psi}_{\rm tot}$, but with the corresponding diagonal entry 
also set to zero. The application of equations~(\ref{eq:C_comb}) and (\ref{eq:D_comb}) 
leads to a final set of consensus constraints, encoding the full information of the 
BAO-only and BAO+RSD analyses. The results corresponding to the intermediate redshift bin are shown by the 
filled contours in Figure~\ref{fig:fullcov_final}, where the reduction in the allowed region of the parameter space with respect to the BAO and RSD results is clear. 

\begin{figure}
\includegraphics[width=0.45\textwidth]{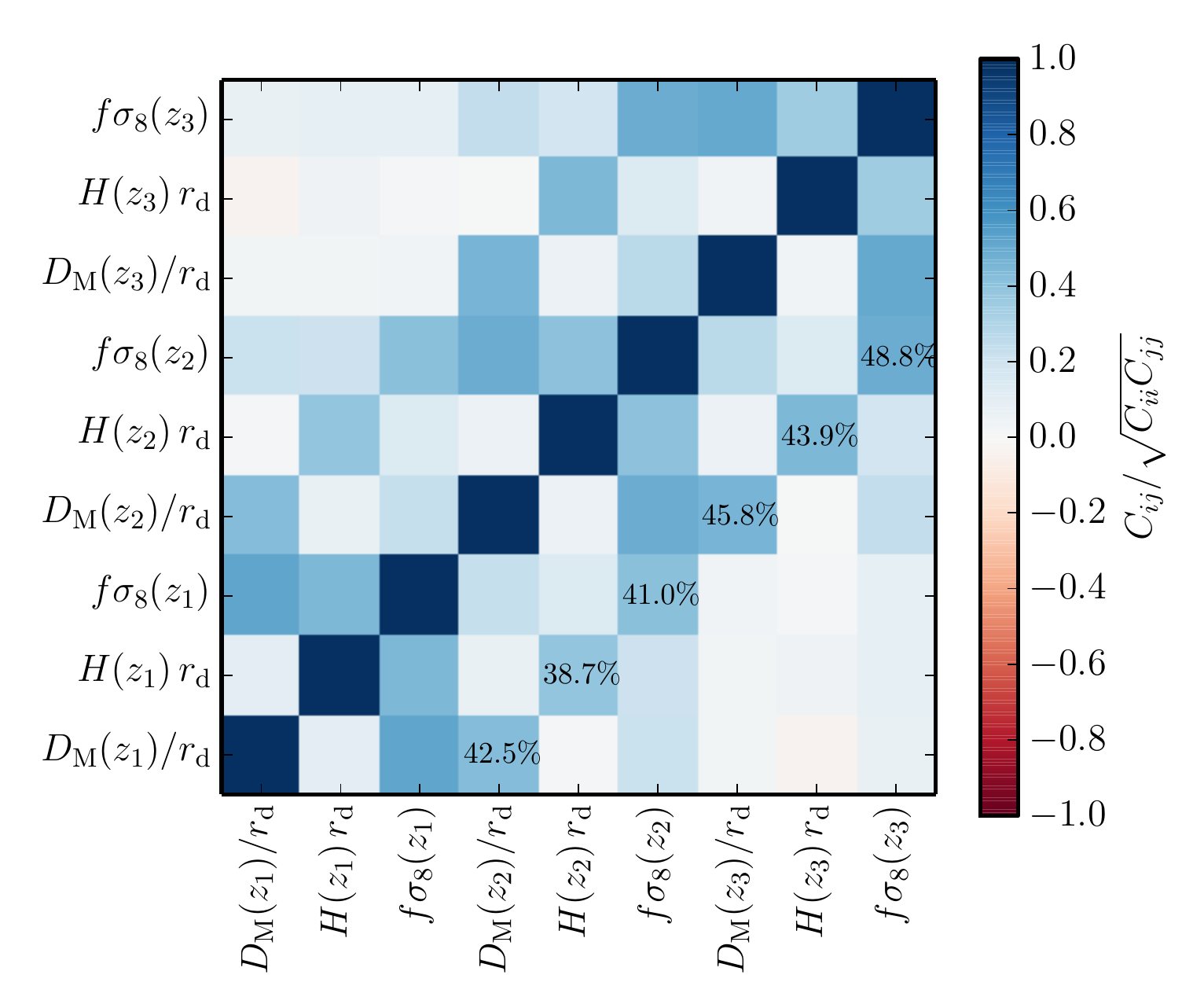}
\caption{
Correlation matrix corresponding to the covariance of the full consensus 
constraints in our three redshift bins recovered from our BOSS mock catalogues.
}
\label{fig:cov_zbins}
\end{figure}

Figure~\ref{fig:cov_zbins} shows the correlation matrix corresponding to the covariance of the 
full consensus constraints recovered from our BOSS mock catalogues in the three redshift bins.
The $3\times3$ blocks along the diagonal correspond to the consensus covariance $\mathbfss{C}_{\rm c}$ 
at each redshift, which show a similar structure. As can be seen from the off-diagonal blocks, the 
consensus constraints of the low- and high-redshift bins are essentially independent, but both 
exhibit a strong correlation with the results of the intermediate one due to the large redshift overlap.
\citet{Acacia2016} use this covariance matrix as the basis of the cosmological implications of the
consensus constraints combining the results of the same BAO-only and BAO+RSD methods studied here.

So far we have assumed that the posterior distributions being combined
are not affected by systematic errors. If they are, these errors will be 
propagated into the consensus values and might lead to biased cosmological constraints.
If the different methods are affected by uncorrelated systematic errors, their impact on the consensus 
results would be reduced. However, if these systematic errors shift the value of a given parameter from
the correct result always in the same direction, this deviation will also be present in the 
combined constraints. The methodologies implemented here to extract cosmological
information from BAO+RSD fits show a small deviation from the correct value of $f\sigma_8(z)$ in
the low and intermediate redshift bins. These shifts are inherited by the final consensus constraints, 
which show a deviation from the true value of 0.59$\sigma$, 0.42$\sigma$ and 0.06$\sigma$ for the 
low- intermediate- and and high-redshift bins, respectively. Although these systematic shifts are smaller
than the statistical uncertainties associated with the consensus constraints, they are taken into account
in \citet{Acacia2016}, where they are used to construct a systematic error budget for the measurements
obtained from the final BOSS galaxy samples.

\section{Conclusions}
\label{sec:conclusions}

We presented a general framework to combine the information of multiple Gaussian posterior
distributions into a set of consensus constraints representing their joint information.
This methodology can be applied to combine the cosmological information 
obtained from different clustering measurements based on the same galaxy sample, which
can often be expressed as Gaussian constraints on a small number of parameters.
The application of this technique requires the knowledge of the full cross-covariances
of the different methods. For clustering measurements, this information can be obtained
using a brute-force approach, applying the same methods being combined to a set of mock 
galaxy catalogues and measuring the correlations between the obtained results.  

We illustrate our technique by applying it to combine the results obtained from different 
BAO-only and BAO+RSD measurements from an ensemble of mock catalogues of the final 
BOSS. The obtained consensus constraints represent a reduction in the 
allowed region of the parameter space with respect to the results of the individual methods. 
This shows the value of using the combination of the results of multiple clustering analyses
as a strategy to maximise the constraining power of galaxy surveys.

In our companion paper \citet{Acacia2016}, the methodology described here is used to
obtain a set of consensus constraints that encode the results obtained by applying the
same methods studied here to the final BOSS galaxy samples. These results are then used
to explore the cosmological implications of the data set in combination with the
information from cosmic microwave background and  Type Ia supernovae data. 

We anticipate that the procedure detailed here can help to optimize the use of the 
cosmological information encoded in future clustering and lensing analyses.

\section*{Acknowledgements}

AGS would like to thank Ximena Mazzalay for her invaluable help in the preparation
of this manuscript. We would like to thank Riccardo Bolze, Daniel Farrow, Jiamin Hou, Martha Lippich
and Francesco Montesano for useful discussions. AGS, JNG and SS-A acknowledge support from the 
Trans-regional Collaborative Research Centre TR33 `The Dark Universe' of the German Research 
Foundation (DFG). CC acknowledges support as a MultiDark Fellow. CC acknowledges support from the 
Spanish MICINNs Consolider-Ingenio 2010 Programme under grant MultiDark CSD2009-00064, MINECO 
Centro de Excelencia Severo Ochoa Programme under grant SEV-2012-0249, and 
grant AYA2014-60641-C2-1-P.

Funding for SDSS-III has been provided by the Alfred P. Sloan Foundation, the Participating
Institutions, the National Science Foundation, and the U.S. Department of Energy. 

SDSS-III is managed by the Astrophysical Research Consortium for the
Participating Institutions of the SDSS-III Collaboration including the
University of Arizona,
the Brazilian Participation Group,
Brookhaven National Laboratory,
University of Cambridge,
Carnegie Mellon University,
University of Florida,
the French Participation Group,
the German Participation Group,
Harvard University,
the Instituto de Astrofisica de Canarias,
the Michigan State/Notre Dame/JINA Participation Group,
Johns Hopkins University,
Lawrence Berkeley National Laboratory,
Max Planck Institute for Astrophysics,
Max Planck Institute for Extraterrestrial Physics,
New Mexico State University,
New York University,
Ohio State University,
Pennsylvania State University,
University of Portsmouth,
Princeton University,
the Spanish Participation Group,
University of Tokyo,
University of Utah,
Vanderbilt University,
University of Virginia,
University of Washington,
and Yale University.

Based on observations obtained with Planck (http://www.esa.int/Planck), an ESA science
mission with instruments and contributions directly funded by ESA Member States, NASA, and Canada.






\end{document}